\documentclass[review]{elsarticle}

\usepackage{lineno,hyperref}
\modulolinenumbers[5]

\journal{Journal of \LaTeX\ Templates}

\begin{document}

\begin{frontmatter}

\title{Association schemes perspective of microbubble cluster in ultrasonic fields}

\author[T1]{S. Behnia\corref{mycorrespondingauthor}}
\cortext[mycorrespondingauthor]{Corresponding author}
\ead{s.behnia@sci.uut.ac.ir}

\author[B1]{M. Yahyavi}
\author[T1]{R. Habibpourbisafar}
\address[T1]{Department of Physics, Urmia University of Technology, Orumieh, Iran.}
\address[B1]{Department of Physics, Bilkent University, 06800 Bilkent, Ankara, Turkey.}

\begin{abstract}
Dynamics of a cluster of chaotic  oscillators on a network are studied using coupled maps. By introducing the association schemes, we obtain coupling strength in the adjacency matrices form, which satisfies Markov matrices property.  We remark that in general, the stability region of the cluster of oscillators at the synchronization state is characterized by Lyapunov exponent which can be defined based on the $N$-coupled map. As a detailed physical example, dynamics of microbubble cluster in an ultrasonic field are studied using coupled maps. Microbubble cluster dynamics have an indicative highly active nonlinear phenomenon, were not easy to be explained. In this paper, a cluster of microbubbles with a thin elastic shell based on the modified Keller-Herring equation in an ultrasonic field is demonstrated in the framework of the globally coupled map. On the other hand, a relation between the microbubble elements is replaced by a relation between the vertices.   Based on this method, the stability region of microbubbles pulsations at complete synchronization state has been obtained analytically. In this way, distances between microbubbles as coupling strength play the crucial role. In the stability region, we thus observe that the problem of study of dynamics of $N$-microbubble oscillators reduce to that  of a  single microbubble. Therefore, the important parameters of the isolated microbubble such as applied pressure, driving frequency and the initial radius have effective behavior on the synchronization state.
\end{abstract}

\begin{keyword}
Globally Coupled Map \sep Lyapunov exponent \sep Associated scheme \sep Bose-Mesner algebra \sep Encapsulated microbubbles.
\end{keyword}

\end{frontmatter}

\section{Introduction}

{{{ Ultrasound contrast agents (UCAs)    are coated microbubbles by a stabilizing shell (polymer, albumin or lipid) which have the medical applications such as diagnostic ultrasound imaging and drug and gene delivery~\cite{kliba,hyum}}}. So far, most of the investigations have been devoted to the dynamics of the single {{microbubble}}. When {{UCAs}} interact with another one, the dynamical behavior of the interaction is completely different from the isolated case. Therefore, a good mathematical modeling of multi-{{microbubble}} dynamics in a cluster becomes  extremely necessary.
The study of radial dynamics of spherical single-bubble was introduced primarily by Rayleigh~\cite{rayleigh} which is formulated as free gas bubble in the incompressible inviscid liquid. Further studies by Plesset and Prosperetti~\cite{plesset1,pless} considered acoustical field for Rayleigh basic equation which called the Rayleigh-Plesset (R-P) model. A complete Rayleigh model is Rayleigh, Plesset,  Noltingk, Neppiras, and Poritsky (RPNNP) equation~\cite{eatoc,noltingk} which include the effects of liquid viscosity, surface tension, and an incident acoustic pressure wave with low acoustic amplitude parameters. Later, Keller-Miksis~\cite{kelle} derived a model for free gas bubble in which the liquid's compressibility can be easily incorporated. The first UCAs model which added a thin viscoelastic albumin-shell and damping coefficient term to the RPNNP equation is  proposed by de jong et al.~\cite{dejonge}.
The shell thickness and rigidity of UCAs in the RPNNP model were also considered by Church~\cite{chur}.
Multi-bubbles were theoretically studied by Takahira by means of the series expansion of the spherical harmonic (Legendre series)~\cite{takahira}. {Doinikov by using the lagrangian formalism and Clebsch-Gordan expansion investigated a mathematical model for collective free gas bubble dynamics in strong ultrasound fields~\cite{doin}}. {{The cluster of microbubbles with a thin encapsulation  added to the Keller-Herring (K-H) equation~\cite{lezzi} have been analyzed by Macdonald and Gomatam~\cite{gomatan}}}. The nonlinear nature of {{above theoretical models}} need specialized tools for analysis due to the fact that linear and analytical solutions are not enough. {{When}} the motion of {{bubbles or UCAs}} gets chaotic,  their theoretically observed behaviors    with chaos theory tools such as bifurcation and Lyapunov diagrams~\cite{parli,behnia1,feng,behnia2} have been studied. For this reason, it is substantial to have appropriate information about the microbubbles dynamics, for finding an acceptable stability region in various applications in industry. Since the K-H model for UCAs is usually not studied  in terms of  $N$ interacting microbubbles, the question arises, what distribution  do the microbubbles correspond to?}

The concept of coupled map lattices (CML) was first suggested by Kaneko~\cite{kaneko1,kaneko2}, which can be demonstrated as an array of smaller finite-dimensional subsystems endowed with local interactions. The CML consists of an array of dynamical elements  which interacts (coupled) with other elements whose values are continuous or discrete in space and time~\cite{waller,kaneko3}. Behaviors discovered in CML have been observed in chemical systems, fluids, electronics, traffic \cite{tadaki1998coupled,tadaki1999noise,yukawa1995coupled}, optical  systems, networks \cite{shinoda2016chaotic}, and as well as in neural dynamics \cite{kaneko2000} and biological, and also in indirect experiments \cite{hagerstrom2012experimental}. An extension of CML, in which each element is connected  with  all other elements is called globally coupled map (GCM)~\cite{just,ming,saito1992behavior}. GCM have diverse applications in a real physical world such as Josephson junction arrays~\cite{hadley} and multi-mode lasers~\cite{laser}.

Significant strides have been made for studying the dynamics of network structures \cite{albert2002statistical,watts1998collective}. T. Gross \textit{et al.}  \cite{gross2009adaptive} specifically focused on the global synchronization among all oscillators. The authors of Ref. \cite{motter2005network} analyzed the design of easily synchronized networks and found synchronizability to be varying.   Synchronization is the most typical collective behavior in complex networks showing  trajectories of each coupled dynamical elements which remains in step with each other during the temporal evolution. Complete synchronization is introduced by Pecora and Caroll in 1990~\cite{pecora}, where by means of synchronization, the state variables of individual systems converge towards each other~\cite{wu}. One of the powerful mathematical technique which has been used by several authors~\cite{belykh,atay,hasler} is the analysis of synchronization corresponding to the associative  relation between the array of coupled oscillators and graph theory.  Bose and Nair~\cite{bosse} in the design of statistical experiments introduced the theory of association schemes. In fact, association schemes as algebraic combinatorics are relations between pairs of elements of a set, which also arise naturally in the theory of permutation groups,  independent of any statistical applications~\cite{bosse,bos}. The governing algebra on the association schemes was formulated by Bose and Mesner~\cite{bose} which is known as the Bose-Mesner algebra. Bose-Mesner algebra of an association scheme is the matrix algebra which generate then by adjacency matrices of the elements of the set. One is lead to ask two questions. The CML gives information of the stability region of which physical quantity? Can one investigate the dynamic behaviors of $N$ interacting microbubble cluster from the CML approach at complete synchronization state?

The answer to the first question depends on the physical context in which the CML is defined. Dynamics of coupled chaotic oscillators on the physical context are studied using coupled maps. The study of CMLs is one significant method to investigate the emergent phenomena, such as synchronization, cooperation, and more,  which may happen in  interacting physical systems. As a physical example, the chaotic nature of the microbubble-microbubble interaction  requires particular tools for resolution, because the analytical and linear solutions are not sufficient.

{Since the K-H model for UCAs is usually not studied  in terms of  $N$ interacting microbubbles~\cite{gomatan,behnia2}. In this paper, to answer the above question, we  employ an association scheme and the Bose-Mesner algebra in order to calculate the stability of $N$-microbubbles in the cluster at complete synchronization state. The coupling strength of the coupled K-H model could generate Markovian matrices and satisfy Markov conditions. In particular, when a cluster of microbubbles is globally synchronized, their
dynamics are reduced  that of a single microbubble. In this case, CML should be relevant for studying the salient behaviors of  microbubble-microbubble interaction. In the present paper, we study complete synchronization of ultrasound contrast agents (UCAs) microbubbles interaction in a cluster. An important advantage of this method is that it is enough to have information only for one microbubble. All the numerical results demonstrate that  association schemes perspective for studying   the radial response of UCA microbubbles is very effective.}

%=============================================================================================================

\section{Definitions: Graph
Theory, Association Schemes and Bose-Mesner Algebra}
In this section, we give some preliminaries such as definitions related to graph theory, association schemes and Bose-Mesner algebra which are used through out the paper~\cite{bailey,brouw}. \\ Graph is a pair $\Omega=(V, E)$, where $V$ is a non-empty set($\Omega$) and $E$ is a subset of $\{(\alpha,\beta): \alpha, \beta \in  V, \alpha \neq \beta\}$ . Elements of the graph are called vertices (V) and edges (E). Two vertices $\alpha,\beta \in V$ are called adjacent if $\{\alpha, \beta\}\in E$.
The adjacency matrix is defined by~\cite{gra,grossk},
$$
(A)_{i,j}=\left\{\begin{array}{cc}
                    1 & \;\;\; \textmd{if} \;\;\; i \sim j \\\\
                     0 & \;\;\;\textmd{otherwise}
                   \end{array}\right.
$$
Obviously, $A$ is symmetric matrix. The valency of a vertex, $i \in V(G)$ is defined as
$$deg(i)\equiv \kappa(i)=\mid\{j \in V(G) : i \sim j \}\mid$$
where $\mid.\mid$ denotes the cardinality (the cardinality of a set is a measure of the number of elements of the set). Let $V$ be a set of vertices and $R_{\alpha}=\{R_{0}, R_{1},... ,R_{d}\}$  be a nonempty set of relations on $V$ which is named  associate class. The pair $\tilde{X}=(V,R_{\alpha})$ is called an association scheme of class $d$ ($d$-class scheme) on $V$ under the following four conditions ~\cite{shimamo,bailey,brouw}:
\begin{enumerate}
  \item  $\{R_{\alpha}\}$ is a part of $V \times V $,
  \item  $R_{0}=\{(i,i):i \in V\}$,
  \item  $R_{\alpha}=R_{\alpha}^{T}$ where $R_{\alpha}^{T}=\{(j,i):(i,j) \in R_{\alpha}\}$,
  \item  For any $(i,j) \in \beta$, the number of $p_{\alpha \beta}^{\gamma}=|\{k \in V:(i,k) \in R_{\alpha} $ and $(k,j) \in R_{\beta}\}|$ depends only on $\alpha,\beta,\gamma$.
\end{enumerate}
where   $\tilde{X}$ is a symmetric and commutative association scheme of class $d$ from conditions (3) and (4), respectively. The elements $i$ and $j$ of $V$ are called $\alpha^{th}$ associates if $(i,j) \in R_{\alpha}$ and $d+1$ is the number of associate classes which is called the rank of the scheme. The intersection numbers of the association scheme are denoted by $p_{\alpha \beta}^{\gamma}$. Indeed,  condition (3) implies that $p_{\alpha \beta}^{0}=0$ if $\alpha \neq \beta$ while $p_{0 \beta}^{\beta}=p_{\alpha 0}^{\alpha}=1$. Also, condition (4) implies that every element of $V$ has $p_{\alpha \alpha}^{0}$ which is defined as
\begin{equation}\label{valencyy}
\kappa_{\alpha}=p_{\alpha \alpha}^{0}
\end{equation}
this is called the valency of $\alpha^{th}$ associates class $(\kappa_{\alpha} \neq 0)$. Relation between the number of vertices (or order of the association scheme) and valency is given by:
\begin{equation}\label{sareza}
N=|V|=\sum_{\alpha=0}^{d}\kappa_{\alpha}
\end{equation}
other definition are given as;
\begin{eqnarray}
 \nonumber
  \sum_{\alpha=0}^{d} A_{\alpha}=J_{N},\quad A_{0}=I_{N},\quad  \\
\label{matrix}
A_{\alpha}=A_{\alpha}^{T} ,\quad A_{\alpha}A_{\beta}=\sum_{\gamma=0}^{d} p_{\alpha \beta}^{\gamma} A_{\beta}.
\end{eqnarray}
$J$ is an $N\times N$ matrix with all-one entries. Also, {{a sequence of matrices}} $A_0, A_1, ..., A_d$  generates a commutative $(d+1)-$dimensional algebra $\mathbf{A}$ of symmetric matrices which is called Bose-Mesner algebra of $\tilde{X}$~\cite{bose}. {{It should be noted that, the matrices $A_{\alpha}$ are commuting and they can be diagonalized simultaneously \cite{marcus}. There exists a matrix ($M$) in such a way that for each, $A\in \mathbf{A}$, $M^{-1}AM$ is a diagonal matrix. Therefore, $\mathbf{A}$ has a second basis $E_{0},..., E_{d}$ \cite{gra,burrow}. These matrices satisfy
\begin{equation}\label{idempotenttt}
E_{0} = \frac{1}{N}J_{N},\quad E_{\alpha}
E_{\beta}=\delta_{\alpha\beta} E_{\alpha},\quad \sum_{\alpha=0}^d
E_{\alpha}=I_{N}
\end{equation}
where $E_{\alpha}, E_{\beta}$, for ($0\leq \alpha,\beta\leq d$) are known as the primitive idempotent of $\tilde{X}$ while matrix $\frac{1}{N}J_{N}$ is a minimal idempotent. If $P$ and $Q$ be the matrices relating to our two bases for $\mathbf{A}$, then~\cite{bailey,brouw}:
\begin{equation}\label{eigen1}
A_{\beta}=\sum_{\alpha=0}^d P(\alpha,\beta)E_{\alpha}, \;\;\;\;\ 0\leq
\beta\leq d,
\end{equation}
\begin{equation}
E_{\beta}=\frac{1}{N}\sum_{\alpha=0}^d Q(\alpha,\beta)A_{\alpha},
\;\;\;\;\ 0\leq \beta\leq d.
\end{equation}
On the other hand, the matrices $P$ and $Q$ satisfy $PQ=QP=NI_{N}$, it also follows that~\cite{bannai,godsill}
\begin{equation}\label{mainly}
 A_{\beta} E_{\alpha}=P(\alpha,\beta)E_{\alpha},
\end{equation}
which shows that $P(\alpha,\beta)$ ($Q(\alpha,\beta)$) is $\alpha$-th eigenvalues ($\alpha$-th dual eigenvalues) of $A_{\beta}$ ($E_{\beta}$) and the columns of $E_{\alpha}$ are the corresponding eigenvectors. Two eigenvalues satisfy
$$
 m_\beta P(\alpha,\beta)=\kappa_{\alpha} Q(\alpha,\beta), \;\;\;\;\ 0\leq
\alpha,\beta\leq d,
$$
where
$$m_{\beta}=Tr(E_{\beta}), \quad \sum_{\alpha=0}^{d} m_{\alpha}=N, \quad m_{0}=1 $$
 Note that, GCM is one of the favorite models in the study of spatially chaotic coupled systems. This model with global coupling strength corresponds to the complete graph~\cite{gup}. Complete graph is a unidirectional graph in which each pair of distinct vertices is connected by a unique edge which is denoted by $K_N$ with $N$ vertices and have $\frac{N(N-1)}{2}$ edges.

\section{Lyapunov Exponent of $N$-coupled Dynamical Systems}
Consider a network of $N$ nodes with $N$ couplings between nodes. Each node of the network can be characterized a dynamical variable $x_{i}$, where $i=1,2,...,N$. Then evolution of coupled dynamical system is written as:
\begin{equation}\label{global}
\dot{x_{i}}(t)= (1-\epsilon)f(x_{i}(t))+\frac{\epsilon}{N}\sum_{j=1}^N f(x_{j}(t))
\end{equation}
Where the above equation is represented GCM model with global coupling $\epsilon$. Function $f$ describes the interaction of individual units, which is assumed to be identical for each pair. By introducing
\begin{equation}\label{mesnerrr}
\mathcal{A}=\sum_{\alpha=0}^d \frac{\epsilon_{\alpha}}{p_{\alpha\alpha}^0} A_{\alpha}
\end{equation}
All adjacency matrix elements could cover the topology of elements of the GCM. $A_\alpha$ is presented as the element of Bose-Mesner algebra and their elements show different coupling topology in graph (It is included $A_0$ and $A_1$ in complete graph).
On the other hand, Markov chains (MCs) $\{X_{t}\}$ on space state $S$ is described by a transition probability. The transition probability in MCs from state $s_{i}$ to $s_{j}$ are denoted by $p_{ij}(t+1,t)=P(X_{t+1}=s_{j}| X_{t}=s_{i})$ where $ s_{i},s_{j} \in S $ ~\cite{isa,seneta}. The transition probabilities of  MCs $\{X_{t}\}$ on state space $S$ are exhibited in the matrices form which are known as transition probability or Markov matrices. Note that the elements of a Markov matrix $P_{t}$ satisfied:
\begin{equation}\label{markovo}
p_{ij}(t+1)\geq 0 ,\quad
\sum_{j \in S}p_{ij}(t+1,t)=1
\end{equation}
Also $\epsilon_{\alpha}$ is the coupling constant in the coupled map lattice, with the condition
\begin{equation}\label{mosavi}
\sum_{\alpha=0}^{d}\epsilon_{\alpha}=1
\end{equation}
could generate the Markov matrix. We can write:
\begin{equation}
\dot{x_{i}}(t)=\sum_{j=1}^N \mathcal{A} f(x_{j}(t)),\quad   j= 1, 2, ..., N.
\end{equation}
One of the significant properties of Markovian matrices is that it should contain the eigenstate $(1,1,1,\ldots)$, where it presents the synchronized state in CML. Now synchronization is one of the invariant manifold of dynamical systems. In order to analyze the stability at the synchronized state by perturbing
\begin{equation}
\delta \dot{x_{i}}(t)= \sum_{j=1}^N \frac{\partial \dot{x_{i}}(t)}{\partial x_{j}(t)}\delta x_{i}(t)
\end{equation}
then, we obtain:
$$\delta \dot{x_{i}}(t)= \sum_{j=1}^{N} \mathcal{A} \frac{\partial f(x_{i}(t))}{\partial x_{j}(t)} \delta x_{i}(t)$$
or
\begin{equation}
\delta \dot{x_{i}}(t)=\mathcal{A} f'(x_{i}(t))\delta x_{i}(t)
\end{equation}
By iterating
$$\delta \dot{x_{i}}(t)=\left(\prod _{m=0}^{t-1} \mathcal{A} f'(x_{i}(m))\right)\delta x_{i}(0)=(\mathcal{A})^m$$
\begin{equation}\label{itera}
 \times \prod _{m=0}^{t-1} f'(x_{i}(m))\delta x_{i}(0)
\end{equation}
Substituting Eq. (\ref{eigen1}) in Eq. (\ref{mesnerrr})
$$\mathcal{A}=\sum_{\alpha=0}^d \frac{\epsilon_{\alpha}}{p_{\alpha\alpha}^0} \sum_{\beta=0}^d P(\beta,\alpha)E_{\beta}$$
this is equivalent to (with respect Eq. (\ref{idempotenttt})):
$$
(\mathcal{A})^m=\sum_{\beta=0}^d \left(\sum_{\alpha=0}^d \frac{\epsilon_{\alpha}}{p_{\alpha\alpha}^0}P(\beta,\alpha)\right)^m E_{\beta}
$$
and so
$$
(\mathcal{A})^m=\sum_{\beta=0}^d \left(\sum_{\alpha=0}^d \frac{\epsilon_{\alpha}}{p_{\alpha\alpha}^0}P(\beta,\alpha)\right)^m E_{\beta}
$$
Now, Eq. (\ref{itera}) can be written as:
$$
\delta \dot{x_{i}}(t)=\sum_{\beta=0}^d \left(\sum_{\alpha=0}^d \frac{\epsilon_{\alpha}}{p_{\alpha\alpha}^0}P(\beta,\alpha)\right)^m$$
$$
 \times E_{\beta} \prod _{m=0}^{t-1} f'(x_{i}(m)) \delta x_{i}(0)
$$
For $\beta=0$ we have
\begin{equation}\label{lyaaa}
\sum_{\alpha=0}^d \frac{\epsilon_{\alpha}}{p_{\alpha\alpha}^0}P(0,\alpha)=1
\end{equation}
that leads us to write
$$\delta \dot{x_{i}}(t)=\prod _{m=0}^{t-1} f'(x_{i}(m)) E_{0} \delta x_{i}(0)+ \sum_{\beta=1}^d \prod _{m=0}^{t-1} f'(x_{i}(m))$$
$$
\times\sum_{\alpha=0}^d \frac{\epsilon_{\alpha}}{p_{\alpha\alpha}^0}P(\beta,\alpha) E_{\beta} \delta x_{i}(0)
$$
where $E_{0} \delta x_{0}$ represent the synchronized state and the other elements $E_{\beta} \delta x_{i}(0)$ are dependent on the transverse state. So the Lyapunov exponent of $N$-coupled dynamical systems is defined as
$$
\Lambda_{\beta}=\lim_{n\longrightarrow \infty} \frac{1}{n} \ln \frac{\|\delta \dot{x_{i}}(t) \|}{\|\delta x_{i}(0)\|}
= \lambda_{f(x_{i}(t))}$$
$$
+ \ln \left(\sum_{\alpha=0}^d \frac{\epsilon_{\alpha}}{p_{\alpha\alpha}^0}P(\beta,\alpha)\right)
$$
$\lambda_{f(x_{i}(t))}$ shows Lyapunov exponent for single dynamical systems \cite{YHV}. For stability of transverse mode it is necessary to have $ \Lambda_{\beta}<0 $  $(\beta=1,2,\cdot \cdot \cdot , d)$:
\begin{equation}
 \left| \sum_{\alpha=0}^{d}\frac{\epsilon_{\alpha}}{p_{\alpha \alpha}^{0}}P(\beta,\alpha) \right| \leq e^{-\lambda_{f(x_{i}(t))}}
\end{equation}
Finally, by separating $\alpha=0$, synchronized state makes the coupling strength of GCM meet following inequality condition:
\begin{equation}\label{lyappa}
1-e^{-\lambda_{f(x_{i}(t))}} \leq \sum_{\alpha=1}^{d}(\kappa_{\alpha}-P(\beta,\alpha))\frac{\epsilon_{\alpha}}{\kappa_{\alpha}}
\leq 1+e^{-\lambda_{f(x_{i}(t))}}
\end{equation}
The stable domain at the complete synchronization state is restricted between the coupling $\epsilon$, association schemes and Bose-Mesner algebra parameters. The number of vertices and associated classes ( Eq. (\ref{sareza}) and Eq. (\ref{eigen1})) have an important role in inequality condition. As mentioned, $\epsilon$ generate Markov matrices which represent the transition probability of MCs random variable $\{x_{t}\}$ (Eq. (\ref{global})). Let vertices of a random graph $(V=1,2,\cdot \cdot \cdot)$ be the space state of a random variable $\{x_{t}\} \in V $. Then $\epsilon$ are Markov matrices on MCs whose elements are transition probabilities from vertex $i$ to $j$, denoted by:
$$
\epsilon(t+1,t)=
=\left(
\begin{array}{ccc}
  \epsilon_{1,1}(t+1,t) &  \cdots & \epsilon_{1,j}(t+1,t) \\
  \epsilon_{2,1}(t+1,t) &  \cdots & \epsilon_{2,j}(t+1,t) \\
  \vdots  & \vdots & \vdots  \\
  \epsilon_{i,1}(t+1,t) &  \cdots & \epsilon_{i,j}(t+1,t)\\
 \end{array}
 \right)
$$
we have considered, $\epsilon(t+1)=\epsilon(t)=\epsilon$. The diagonal elements of Markov matrix is $\epsilon_{i,i}=1-\epsilon$ (for associate class $\alpha=0$) and other matrix elements are $\epsilon_{i,j}=\frac{\epsilon}{\kappa_{\alpha}}$  $(\alpha=1,2,...,d)$.

\section{Theoretical Model for a Cluster of Microbubbles}
A cluster of $N$ interacting microbubbles on the basis of the general K-H equation~\cite{lezzi} for $i^{th}$ microbubble is given by:

$$\left[1-({b}+1)\frac{\dot{R_{i}}}{c}\right]R_{i}\ddot{R_{i}}+\frac{3}{2}\left[1-(3 {b}+1)\frac{\dot{R_{i}}}{3c}\right]\dot{R_{i}}^{2}$$
\begin{equation}\label{keller}
=\frac{1}{\rho}\left[1+(1-{b})\frac{\dot{R_{i}}}{c}+\frac{R_{i}}{c}\frac{d}{dt}\right]\left(P_{i}(R,\dot{R})-P_{ext,i}(t)\right)
\end{equation}
where
\begin{equation}\label{akhatovv}
P_{ext,i}(t)=P_{ac}\sin (2\pi \nu t)+\sum_{{j=1, j\neq i }}^{N} \frac{\rho}{D_{ij}}\frac{d}{dt}\left(R_{j}^{2}\dot{R_{j}}\right)
\end{equation}

with an explicit expression for $P_{i}(R,\dot{R})$ which developed by Morgan et al.~\cite{INTRODUCTION--Morgan}, allowing for the encapsulating shell, is defined as
\begin{eqnarray}\label{M-P}
 \nonumber
  P_i(R,\dot{R})=\left(P_{0}
+\frac{2(\sigma+\chi)}{R_{i0}}\right)(\frac{R_{i0}}{R_{i}})^{3\Gamma}-\frac{4\mu\dot{R}_{i}}{R_{i}}-\frac{2\sigma}{R_{i}}\\
\label{Pi}-\frac{2\chi}{R_{i}}(\frac{R_{i0}}{R_{i}})^{2}-12\mu_{sh}\varepsilon \frac{\dot{R}_{i}}{R_{i}(R_{i}-\varepsilon)}-P_{0}
\end{eqnarray}
In Eq.(\ref{M-P}), replacing subscripts $i$ and $j$ yields the equation for microbubble $j$. Where
$\ddot{R_i}$ is UCAs wall acceleration,
$\dot{R_i}$ is UCAs wall velocity,
$R_i$ is the time-dependent UCAs radius,
$R_{i0}$ is the initial radius for microbubble $i$,
$N$ is the number of microbubbles in cluster,
$D_{ij}$ is the distance between microbubble $i$ and $j$,
$\mu_{sh}$ is the viscosity of the shell,
$\varepsilon$ is the shell thickness,
$\mu$ is the viscosity of the liquid,
$c$ is sound velocity in liquid,
$\Gamma$ is polytropic exponent for UCA gas,
$\chi$ is the shell elasticity,
$\sigma$ is the surface tension,
$\rho$ is the density of the liquid surrounding of the microbubbles,  $P_{0}$ is the ambient static pressure,  $P_{ac}$ is the driving external pressure, and
$\nu$ is frequency.

A Keller-type and Herring-type equation is obtained for {{$b=0$ and $b=1$}}, respectively. Pressures acting on the microbubbles are not equal to the external driving pressure because the amplitude of the pressure waves radiated by the neighboring microbubbles is no longer negligible~\cite{mett}.
In order to perform an analysis of Bose-Mesner perspective, it is convenient to transform the second order differential equation into an autonomous system of first-order differential equations. Hence, we consider
\begin{equation}
\dot{R}_{i}=U_{i}, \quad \dot{\theta}=\nu
\end{equation}
Then, for {$i^{th}$} element (Eq. (\ref{keller})):
$$
\frac{d}{dt}\left(R_{i}^{2}U_{i}\right)=\frac{\left(\xi R_{i}U_{i}-2\zeta \right)^{2}}{\xi^{2}\left(\zeta -R_{i}\left(1+\xi U_{i}\right)\right)}$$
$$\times \left(\zeta(\beta +\frac{2}{R_{i}})-\frac{1}{2}\left[1+\frac{\left(\beta -1\right)\left(R_{i}\xi U_{i}-2\zeta\right)}{\left(\beta +1\right)R_{i}}\right]\right)$$
$$
+\frac{R_{i}U_{i}}{\xi}-\frac{R_{i}^{2}U_{i}\left(1+\xi U_{i}\right)}{\xi \left(R_{i}\left(1+\xi U_{i}\right)-\zeta\right)}+\left(\zeta U_{i}+2R_{i}U_{i}\right)$$
$$
+\frac{2R_{i}}{\xi^{2}}\left(\frac{U_{i}-1}{\xi}-\frac{2\zeta}{R_{i}}-1\right)^{2}$$
$$
+\frac{R_{i}}{\rho \left(\zeta-R_{i}\left(1+\xi U_{i}\right)\right)} \left(R_{i}+\frac{\beta+3\Gamma-1}{1+\beta}\left(R_{i}\xi U_{i}-2\zeta\right)\right)$$
$$
\times \left(p_{0}+\frac{2\sigma}{R_{i0}}\right)\left(\frac{R_{i0}}{R_{i}}\right)
+\frac{2\pi \nu \cos(2\pi \theta)R_{i}^{3}P_{ac}}{\rho c \left(\zeta-R_{i}\left(1+\xi U_{i}\right)\right)}$$
$$
+\frac{R_{i}}{\rho \left(R_{i}\left(1+\xi U_{i}\right)-\zeta \right)}\left(R_{i}-\frac{1-\beta}{1+\beta}\left( R_{i}+R_{i}\xi U_{i}-2\zeta\right)\right)$$
$$
\times \left(P_{0}+P_{ac}\sin(2\pi \theta)\right)-\frac{\left(\frac{2\sigma \beta}{\beta+1}-4\mu\right)}{\rho \xi \left(\zeta-R_{i}\left(1+\xi U_{i}\right)\right)}$$
\begin{equation}\label{kolli}
\times \left(R_{i}\xi U_{i}-2\zeta\ -2\sigma R_{i}\right)-\sum_{{
   j=1,
   j\neq i
  }}^{N} \frac{1}{D_{ij}}\frac{d}{dt}\left(R_{j}^{2}U_{j}\right)
\end{equation}
where $\xi=-\frac{\beta+1}{c}$, $\zeta=\frac{4\mu}{\rho c}$.  {Moreover, by investigation of dynamical behaviors of single targeted microbubble by using  Lyapunov exponent diagrams versus control parameters, the most effective parameter in order to control interacting microbubbles cluster become available. Based on plotting Lyapunov exponent diagrams for two important parameter, their intrinsic  behaviors reveals on stable synchronization are discussed in the following sections.}

{
\subsection{Isolated UCA microbubble}
\label{Bubble-Bubble Interaction}
In this study, the radial dynamics of  UCA microbubbles are modelled by using  the general K-H equation \cite{INTRODUCTION-31}, derived by Prosperetti and Lezzi \cite{lezzi}. This justified equation also explains the effect of variation of the shell on the UCA behavior. This class of models contains the elastic shell which makes the microbubble to represent  nonlinear acoustic properties~\cite{gomatan}. The K-H equation for an UCA with a thin elastic shell is  given by the following equation:
$$\left[1-(1+b)\frac{\dot{R}} {c}\right]R{\ddot{R}}+\frac{3} {2}
\left[1-(3b+1)\frac{\dot{R}} {3c}\right]{\dot{R}^2}$$
\begin{equation}\label{ll}
=\frac{1}
\rho \left[1+(1-b)\frac{\dot{R}} {c}+\frac{{R}}
{c}\frac{d}{dt}\right]\times [P(R,\dot{R})-P_0-P_{ac}\sin (2\pi \nu t)]
 \end{equation}
with an explicit expression for $P(R,\dot{R})$   developed by Morgan et al. \cite{INTRODUCTION--Morgan}, allowing for the encapsulating shell, is defined by Eq. \ref{M-P}. In this equation $R=R(t)$ is the UCA's radius, $R_0$ is the initial radius and $P(R,\dot{R})$ demonstrates the pressure on the liquid aspect of the interface for an isolated UCA microbubble~\cite{gomatan,INTRODUCTION--Morgan}. The expression $P_0+P_{ac}\sin (2\pi \nu t)$ shows the pressure in the liquid far from the microbubble, with $P_{0}$ being the ambient static pressure and $P_{ac}\sin (2\pi \nu t)$ the acoustic forcing term. The model was solved for isolated microbubble  using the values of the physical constants  represented in Table. \ref{Table1} for Albunex \cite{gomatan,dejonge,chur}.
\begin{table}
\caption{Constant parameters used in the general Keller-Herring
equation for an ultrasound contrast agent microbubble (for a
bubble/water system at $20^\circ$ $C$).}\label{Table1}
\begin{center}
\begin{tabular}{llll}
Symbol & Description &Units&Value \\
\hline \noalign{\smallskip}
 $\mu$    &   Viscosity        & $ \textmd{Ns/m}^2$    & $0.001$ \\
 $\sigma$ &   Surface tension  &  $ \textmd{N/m}$      &$0.072$ \\
 $c$      &   Sound velocity   & $\textmd{m/s}$        & $1481$ \\
 $p_0$    &   Static ambient pressure   &  $\textmd{N/m}^2$ & $1.01 \times 10^5$ \\
 $\rho$   &   Liquid density   &$\textmd{kg/m}^3$ &$998$ \\
 $\chi$   & Shell elasticity   &  $\textmd{N/m},$ &$8$ \\
 $\varepsilon$& Shell thickness   &  $\textmd{m}$  & $15\times10^{-9}$  \\
 $\mu_{sh}$& Shell viscosity        &    $\textmd{Pa}\;\textmd{s}$      &                $1.77$\\
  $\Gamma$& Polytropic exponent    &  &$1.33$  \\
\end{tabular}
\end{center}
\end{table}
Here, we explained the dynamics of only one UCA microbubble in ultrasonic field by using standard methods of nonlinear dynamics and theory of deterministic chaos. We did all these through plotting and evaluating the Lyapunov exponent spectra. The maximum Lyapunov exponents, approximated computationally for a wide range of injection values, clearly indicates the chaotic behavior of microbubble interaction dynamics.\\
\textbf{Effect of acoustic pressure.}\\
We examine the stability of  an isolated  microbubble~\cite{dejonge,chur} in ultrasonic field  by considering the driving pressure amplitude and the initial radius of the microbubble. Radial motion of single UCA microbubble dynamics is investigated versus a prominent domain of acoustic pressure from $10$ kPa to $2$ MPa. Fig. \ref{FigB1} shows the Lyapunov exponent spectra of the normalized UCA microbubble radius when acoustic pressure of the UCA is taken as the control parameter for several values of frequency and initial radii of the UCA, where stable and chaotic pulsations can be observed in each. It is clear that the chaotic oscillations of UCA appeared by increasing the values of applied pressure,    the microbubble demonstrates more chaotic oscillations as the pressure is intensifying (adapted from ref. \cite{behnia1}).\\
\begin{figure}
\includegraphics[scale=0.59]{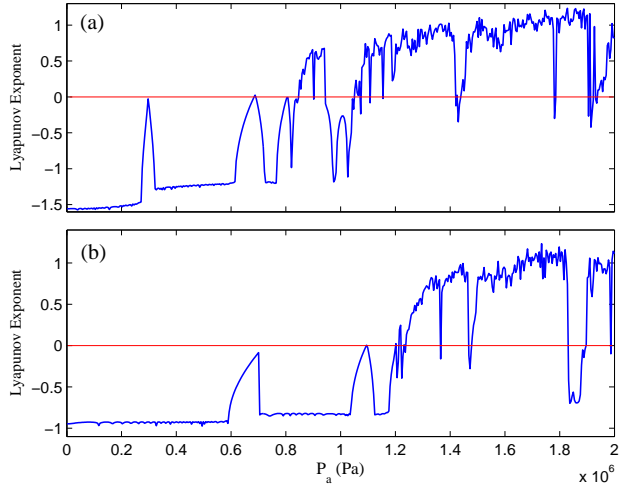}
\caption{ Lyapuonov
spectrum of the normalized microbubble radius versus pressure
when (a) $R_0= 5$ $\mu$m, and $f=1.5$ MHz (b) $R_0= 6$ $\mu$m, and $f=2.5$ MHz.}\label{FigB1}
\end{figure}
\begin{figure}
\includegraphics[scale=0.59]{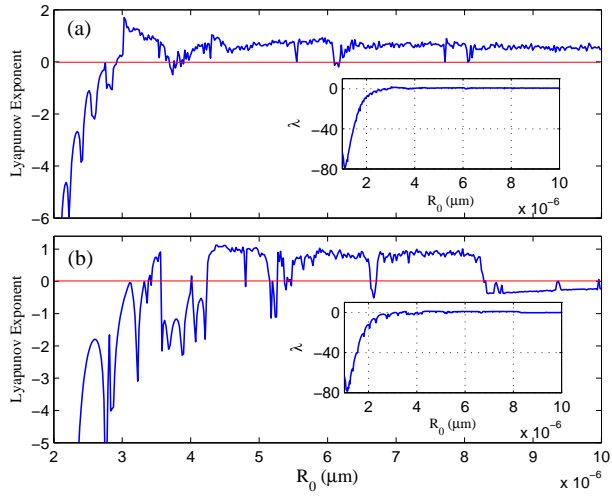}
\caption{Lyapuonov
spectrum of the normalized microbubble radius versus initial radius
when (a) $P_{ac}= 1.5$ MPa, and $\nu=1$ MHz (b) $P_{ac}= 1.5$ MPa, and $\nu=2$ MHz.}\label{FigB2}
\end{figure}
\textbf{Effect of initial microbubble radius.}\\
Also, we examine the variation of initial radius on microbubble dynamics by considering the initial radius as a control parameter. In Fig. \ref{FigB2} we can see the motion of microbubble in ultrasonic fields when driven by the pressure amplitude  limited to 1.5 MPa and the applied frequency is 1 and 2 MHz, respectively. It is observed that the microbubble behavior is stable for high values of frequency. It can be understood from the results are that the motions of microbubble can be chaotic or stable in particular ranges. The results are in agreement with the previous studies clearly highlighting that microbubbles are dependent on the driving frequency variations~\cite{gomatan,SECTION3-2,SECTION3-3}. Most of the results demonstrate the uncontrollable and chaotic motion in UCA microbubble dynamics (see Table \ref{Table2} for more details of the different parameters). In dissimilar situations and values for controlling parameters such as: pressure, frequency, shell thickness and the initial microbubble radius, microbubble  shows various motions and
oscillations by themselves and in addition they change their motion  from one type to another. Having a proper knowledge about microbubbles motion, is the chief motivation in controlling chaotic behavior of the microbubble and using them. \\So according to the light of the above discussion, it can be stated that acoustic forcing term demonstrates its influence on the microbubble dynamics (see Table \ref{Table2}). As a result, acoustic pressure amplitude and frequency of the acoustic are two important factors in the stability of radial pulsation of the microbubble dynamics.}
\begin{table}
\caption{Domain of values of parameters    that lead to chaotic oscillations in the general
K-H equation. (for a bubble/water system at $20^\circ$ C).  All other physical parameters were kept constant at values given in Table \ref{Table1}.}\label{Table2}
\begin{center}
\begin{tabular}{llll}
Effect & Domain of Chaotic Oscillations    \\
\hline \noalign{\smallskip}
                    &    $\nu=1$,       $\varepsilon=15$,     $R_{0}=10$              $\;\;\;\;P_{ac}>0.5$   \\
                    &   $\nu=2$,        $\varepsilon=15$,     $R_{0}=7$                 $\;\;\;\;\;\;P_{ac}>1$     \\
Pressure (MPa)      &   $\nu=3$,        $\varepsilon=15$,     $R_{0}=5$                $\;\;\;\;\;\;P_{ac}>1.5$   \\
                    &   $\nu=3$,        $\varepsilon=15$,     $R_{0}=7$                $\;\;\;\;\;\;P_{ac}>3$     \\
                    &   $\nu=1.5$,      $\varepsilon=15$,     $R_{0}=5$                 $\;\;\; P_{ac}>0.4$   \\  \noalign{\smallskip} \hline \noalign{\smallskip}

                         &  $P_{ac}=1.2$,   $\nu=2$,      $\varepsilon=15$    $\;\;4<R_{0}<7$    \\
                         &  $P_{ac}=1.5$,   $\nu=2$,      $\varepsilon=15$    $\;\;4<R_{0}<8$    \\
Initial radius ($\mu$m)  &  $P_{ac}=2$,     $\nu=2$,      $\varepsilon=15$    $\;\;\;\;\;4<R_{0}<9$    \\
                         &  $P_{ac}=3$,     $\nu=1$,      $\varepsilon=15$    $\;\;\;\;\;1<R_{0}<10$   \\
                         &  $P_{ac}=1$,     $\nu=2$,      $\varepsilon=15$    $\;\;\;\;\;6<R_{0}<7$    \\  \noalign{\smallskip} \hline \noalign{\smallskip}

                       &$P_{ac}=1$,       $\nu=2$,        $R_{0}=3$    $\;\;\;\;\; \varepsilon<5$  \\
                       &$P_{ac}=1$,       $\nu=1$,        $R_{0}=4$    $\;\;\;\;\; \varepsilon<14$ \\
 Shell thickness  (nm) &$P_{ac}=2$,       $\nu=2$,        $R_{0}=6$    $\;\;\;\;\; \varepsilon<15$ \\
                       &$P_{ac}=0.5$,     $\nu=1$,        $R_{0}=6$    $\;\; \varepsilon<6$  \\
                       &$P_{ac}=2$,       $\nu=1$,        $R_{0}=5$    $\;\;\;\;\; \varepsilon<18$ \\

\noalign{\smallskip}

\end{tabular}
\end{center}
\end{table}

\subsection{Microbubbles synchronization state}

We remark that in general, the synchronization state is a physically well-defined observable, which is thought as not  corresponding to an operator acting on the system.  When the underlying microbubbles synchronization state is well defined the associated Lyapunov exponent can be defined basis  on
the $N$-coupled map. However, in practice,  only a finite
number of microbubbles are available. Below, we will calculate the microbubbles synchronization state up to third  order (three microbubbles) in real physical models. The coupling strength of the introduced bubble cluster with three elements (Eq. (\ref{kolli})),  arranged at the vertices of an equilateral triangle ($D_{12}=D_{13}=D_{23}=D$), could be represented by:
$$\epsilon=\left(
    \begin{array}{ccc}
      \beta' & \alpha' & \alpha' \\
      \alpha' & \beta' & \alpha' \\
      \alpha' & \alpha' & \beta' \\
    \end{array}
  \right)
$$
where
\begin{equation}
\alpha'=\frac{D}{2-D^{2}-D},\quad  \beta'=1-\alpha'.
\end{equation}
In order to perform the associated scheme, in the first step, one should find the order of the association scheme $N=|V|=3$ (Eq. (\ref{mesnerrr})). By considering   Eq. (\ref{sareza}), valency of the $i^{th}$ associates is found as:
\begin{equation}
\kappa_{0}=1, \quad \kappa_{1}=2
\end{equation}
also by considering   Eq. (\ref{valencyy}) and Eq. (\ref{mesnerrr})
\begin{equation}
\epsilon_{0}=\beta', \quad \epsilon_{1}=\frac{\alpha'}{2}
\end{equation}
finally
$$
\mathcal{A}=\left(
   \begin{array}{ccc}
    \beta'& \frac{\alpha'}{2} &\frac{\alpha'}{2}\\
     \frac{\alpha'}{2}& \beta'& \frac{\alpha'}{2} \\
    \frac{\alpha'}{2}& \frac{\alpha'}{2} & \beta' \\
   \end{array}
 \right)
$$
or
\begin{equation}\label{markovooo}
\mathcal{A}=\beta' \left(
                                                                                  \begin{array}{ccc}
                                                                                    1 & 0 & 0 \\
                                                                                    0 & 1 & 0 \\
                                                                                    0 & 0 & 1 \\
                                                                                  \end{array}
                                                                                \right)
+\frac{\alpha'}{2}\left(
                                                                                        \begin{array}{ccc}
                                                                                          0 & 1 & 1 \\
                                                                                          1 & 0 & 1 \\
                                                                                         1 & 1 & 0 \\
                                                                                        \end{array}
                                                                                      \right)
\end{equation}
generates the following adjacency matrices having rows and columns corresponding to the vertices of the complete graph:
\begin{equation}
  A_{0}=\left(
          \begin{array}{ccc}
            1 & 0 & 0 \\
            0 & 1 & 0 \\
            0 & 0 & 1 \\
          \end{array}
        \right)=I_{3} , \quad
A_{1}=\left(
        \begin{array}{ccc}
          0 & 1 & 1 \\
          1 & 0 & 1 \\
          1 & 1 & 0 \\
        \end{array}
      \right).
\end{equation}
where it is satisfied in Eq. (\ref{mosavi}). One can show that the corresponding minimal idempotent (Eq. (\ref{idempotenttt})) is
 \begin{equation}
  E_{0}=\frac{1}{3}\left(
          \begin{array}{ccc}
            1 & 1 & 1 \\
           1 & 1 & 1 \\
            1 & 1 & 1 \\
          \end{array}
        \right)
 , \quad
      E_{1}=\frac{1}{3}\left(
              \begin{array}{ccc}
               2 & -1 &-1\\
                -1 & 2 & -1 \\
                -1 & -1 & 2 \\
              \end{array}
            \right).
\end{equation}
in order to determine the stable region ( Eq. (\ref{lyappa})), we need to find eigenvalue of associated schemes $P(\alpha,\beta)$ (Eq. (\ref{mainly})):
\begin{equation}
 P(1,0)=1 ,\quad P(1,1)=-1
\end{equation}
Finally, we have:
\begin{equation}\label{region}
 \frac{2}{3}\left(1-e^{-\lambda_{f(x_{i})}}\right)\leq\ \frac{D}{2\left(2-D^{2}-D\right)} \leq \frac{2}{3}\left(1+e^{-\lambda_{f(x_{i})}}\right)
\end{equation}
Equation~(\ref{region}) describes an inequality condition which is determined by a single microbubble Lyapunov exponent and distances. In fact, the above equation represented transition probability from one site (vertex) to another site in  the microbubble graph. UCAs properties could influence this probability and therefore the synchronization state as well. For the cluster with $N$ elements, it is simple to generate the general form of Eq. (\ref{region}).

 {On the other hand, permissible amounts of distances between microbubbles in order for  UCAs cluster to become in complete synchronization state, it is obtained analytically by using  $\lambda$ over control parameters of single UCAs. It should be noted that, $\lambda$ has three value, when $\lambda<0$, single targeted microbubble is  in stability region and has a periodic behavior, if $\lambda=0$, system has quasi-periodic behavior which shows microbubble tending to transit the chaotic region. For $\lambda>0$, single encapsulated microbubble represent its chaotic nature.} Actually, study of the dynamics of $N$-micobubbles cluster to detect the stable domain at synchronized state is reduced to the study of a single microbubble dynamics (Eq. (\ref{lyappa} ). On the other hand, it is directly dependent on the distance between them.  { We can generalize the  adjacency matrices of Eq. \ref{markovooo} for $N$-microubbles arranged at $N$ vertex of the complete graph,  given as
\begin{equation}\label{r}
\mathcal{A}=\beta' I_N+\frac{\alpha'}{2} (J_N-I_N)
\end{equation}
where $J$  is ones matrix and $I$ is the identity matrix. The rows and columns of these matrices  corresponds to the vertices of the complete graph. The  adjacency matrices of Eq. \ref{r} is completely dependent on $\alpha'$ and $\beta'$ (or the distance between the microbubbles), which we can easily be extended to $N$-microbubble.  For the cluster with $N$ elements it is simple to generate the general form of Eq. (\ref{region}) such as:
$$\frac{N-1}{N}\left(1-e^{-\lambda_{f(x_{i})}}\right)\leq\ \frac{D}{\left(N-1\right)\left(N-1-D^{2}-D\right)}$$
\begin{equation}
\leq \frac{N-1}{N}\left(1+e^{-\lambda_{f(x_{i})}}\right)
\end{equation}}
In such cases, by changing the distances between  microbubble, we achieve variable synchronization condition.  The inequality condition shows permissible values of the distance between microbubbles, suggesting these values are in the stability region. Recent studies  have demonstrated that when inter-bubble distances in a cluster are small, the effect  of coupling between the bubbles can be significant~\cite{manasseh}.  As lifetimes or sizes of microbubbles are different and the distribution of cavitation bubbles is inhomogeneous, the description of the motion of microbubbles become very complicated. Future work could be concentrated on this function.
 {
\section{RESULTS $\&$ Discussion}
The behavior of single microbubble in a cluster is dependent on the condition of host media, applied pressure ($P_{ac}$), driving frequency($\nu$), initial radius($R_{0}$) and viscosity($\mu$). A full discussion of the effects of these parameters on single microbubble are presented in previous section (See Fig. \ref{FigB1}  and Fig. \ref{FigB2}). Studying the dynamics of $N$ bubble cluster for detecting the stable domain at synchronized state is reduced to the study of a single bubble dynamics (Eq. (\ref{lyappa})). At same time, it is found to be directly dependent on distances between them (Eq. (\ref{region})). In addition, it  provides a variable synchronization condition. The inequality condition shows permissible values of distance between microbubbles, and these values they are in the stability region. Recent studies  have demonstrated that when inter-bubble distances in a cluster are small, the effects of coupling between the bubbles can be significant~\cite{manasseh}.
\begin{figure}
\includegraphics[scale=0.54]{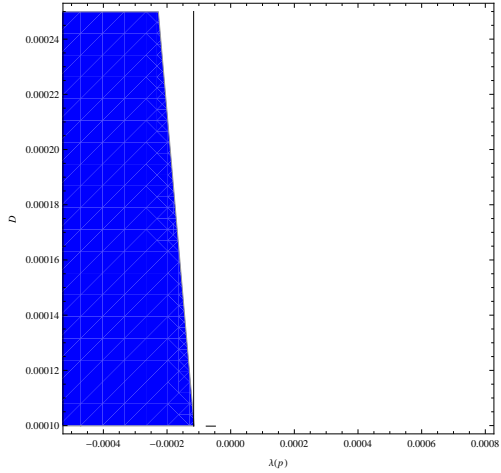}
\caption{Stability region of three microbubbles: Variation of distances between of microbubbles respect to single UCA Lyapunov exponent when $\lambda_{f(x_{i})}$ is a function of applied pressure. }\label{FigS1}
\end{figure}
\begin{figure}
\includegraphics[scale=0.54]{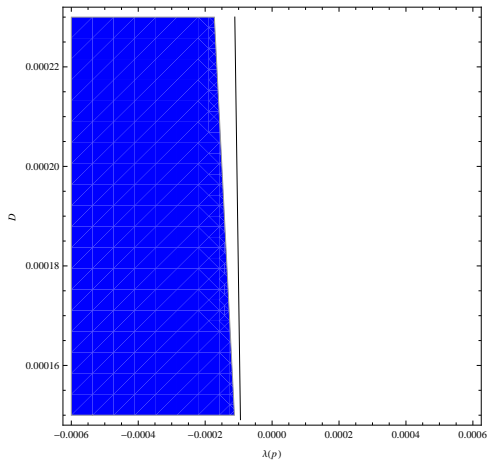}
\caption{Stability region of three microbubbles: Variation of distances between of microbubbles respect to single UCA Lyapunov exponent when $\lambda_{f(x_{i})}$ is a function of driving frequency.}\label{FigS2}
\end{figure}
By varying the pressure, permissible value for synchronization with respect to the distances is restricted in the region  $100\leq D(\mu m)\leq 250$.  Critical value of pressure for synchronization is founded as ($\lambda_{P_{ac}}\sim -0.0002$)(See Fig \ref{FigS1}). Figure \ref{FigS2} depicts,  if $\lambda_{f(x_{i})}$ is a function of driving frequency, then the permissible value for synchronization with respect to the distances is $160\leq D(\mu m)\leq 200$. Bubbles are in the stability region when the frequency varies in the range $500 \leq\nu(MHz)\leq 2000$. Frequency could be defined at the specified value ($\lambda(f)\sim -0.0001$) as a limit for synchronization, which is determined by a tangent  in figure \ref{FigS2}. As lifetimes or sizes of bubbles are different and the distribution of cavitation bubbles is inhomogeneous, the description of the motion of bubbles is very complicated. Future work could be concentrated on this function.}

 \section{Conclusion}

This paper explained the dynamics driven interaction between the cluster of chaotic  oscillators on a network by using the techniques of coupled maps. Here, in this technique, we employed  association scheme and the Bose-Mesner algebra in order to calculate the stability of $N$-oscillators in a cluster at complete synchronization state. Moreover, taking the microbubble-microbubble interaction  as an example we have shown that synchronization condition of $N$-microbubbles in a regular chaotic networks composed of identical elements with symmetric coupling strength can be considered as adjacency matrices. On the other hand, several mathematical models, identify bubbles oscillations in a cluster, such as linear theory \cite{ida2002characteristic,zhang1997momentum} or self-consistent oscillator model \cite{feuillade1995scattering,ye1997sound}.
When the number of bubbles in a cluster is increased, a significant error between experimental and theoretical results appear. In the previous studies \cite{cheng2013interaction}, the applied procedure can facilitate the understanding of cluster formation from the ultrasound echoes and the stable behavior of UCAs network at synchronous mode. When single UCAs in an ultrasonic field demonstrate chaotic behavior $\lambda>0$, we perceive that the acceptable bounds on stable synchronous mode become smaller than that  in the stable or quasi-periodic case. The distances between interacting UCA clusters mainly acquire significance in their synchronization states, which shows that the influence of coupling between microbubbles is always significant, or at least are no longer negligible~\cite{feng}.

\section{Acknowledgement}
Author M.Y would like to thanks Deepak Kumar Singh for  insightful discussions and his help in preparing the manuscript.

\end{document}